\documentclass[]{interact}

\usepackage{epstopdf}
\usepackage{subfigure}
\usepackage[numbers,sort&compress]{natbib}
\bibpunct[, ]{[}{]}{,}{n}{,}{,}
\usepackage{color}
\usepackage{soul}

\begin{document}

\title{A comparison of machine learning techniques for taxonomic classification of teeth from the Family Bovidae}

\author{
\name{G. J. Matthews\textsuperscript{a}\thanks{CONTACT G.J. Matthews. Email: gmatthews1@luc.edu}, J.K Brophy\textsuperscript{b}, M. P. Luetkemeier\textsuperscript{a}, H. Gu\textsuperscript{a}, and G.K. Thiruvathukal\textsuperscript{ac}}
\affil{\textsuperscript{a}Loyola University Chicago,
Chicago, IL
USA;\\
\textsuperscript{b}Louisiana State University,
Baton Rouge, LA
USA;\\
\textsuperscript{c}Argonne National Laboratory,
Argonne, IL
USA.}
}

\maketitle

\begin{abstract}
This study explores the performance of modern, accurate machine learning algorithms on the classification of fossil teeth in the Family Bovidae. Isolated bovid teeth are typically the most common fossils found in southern Africa and they often constitute the basis for paleoenvironmental reconstructions. Taxonomic identification of fossil bovid teeth, however, is often imprecise and subjective. Using modern teeth with known taxons, machine learning algorithms can be trained to classify fossils.  Previous work by Brophy \cite{BrophyEtAl2014} uses elliptical Fourier analysis of the form (size and shape) of the outline of the occlusal surface of each tooth as features in a linear discriminant analysis framework. This manuscript expands on that previous work by exploring how different machine learning approaches classify the teeth and testing which technique is best for classification.  Five different machine learning techniques including linear discriminant analysis, neural networks, nuclear penalized multinomial regression, random forests, and support vector machines were used to estimate these models. Support vector machines and random forests perform the best in terms of both log-loss and misclassification rate; both of these methods are improvements over linear discriminant analysis. With the identification and application of these superior methods, bovid teeth can be classified with higher accuracy.  
\end{abstract}

\begin{keywords}
Classification, Machine Learning, Anthropology
\end{keywords}

\section{Introduction}
Paleoenvironmental reconstruction plays an important role in the study of human ancestors', or hominin, evolution in South Africa. In order to reconstruct past environments, researchers typically rely on the animals that are found assocated with hominins. Animals in the Family Bovidae such as antelopes and buffaloes are of particular interest because they have strict ecological tendencies and are useful for reconstructing the past environments.  However, overlap in the size and shape of teeth in the same taxonomic group can make classifying fossilized remains difficult. This difficulty is exacerbated by the often fragmentary nature of the fossils. In addition, the most common bovid fossils typically recovered from South African sites are isolated teeth. As a result, accurate classification of these fossilized teeth is of tremendous importance in reconstructing paleoenvironments.

Traditionally, researchers rely on fossil and modern comparative collections to identify fossil bovids (\cite{Vrba1976}; \cite{AdamsConroy2005}; \cite{deRuiter2008}) though biasing factors such as age and sex can cause overlap in the size/shape of teeth and can make this approach potentially subjective. A large female of one group can overlap with the size of a small male of another group. Also, as a bovid ages, its teeth wear down. The upper and lower first molars are slightly V-shaped in profile so as they wear down, they appear smaller. These issues coupled with differences in fossil preservation and confidence in identifications can lead to interobserver error. \cite{BrophyEtAl2014} developed a methodology to objectively classify fossil bovid teeth by quantifying the outline of their occlusal surface.

In the aforementioned study, linear discriminant analysis (LDA) was tested to see if you could distinguish between known taxonomic groups of bovids.  Analyses were performed on three maxillary (i.e. UM1, UM2, UM3) and three mandibular molars (i.e. LM1, LM2, LM3) from twenty, known, extant species across seven taxonomic tribes \cite{Brophy2011}.  These prior results indicate that the form (size and shape) of occlusal surface outlines of teeth from the Family Bovidae reliably differentiates between bovid species in the same tribe.  Due to the success of classifying known, extant bovid teeth, the methodology was applied to identifying unknown, fossil teeth.  The amplitudes of each fossil tooth were compared to the entire modern reference sample of the same tooth type (e.g. LM2) using LDA to predict first to what tribe and then to what species each fossil most appropriately belonged.


While Brophy \cite{Brophy2011} demonstrates that the outline of the occlusal surface can be used to reliably discriminate between bovid taxa using LDA, more advanced algorithms exist that allow for models that are more flexible than a LDA and will likely lead to more accurate classification \cite{TongEtAl2011, HuangEtAl2009, Dixon2009, Brown2012, Tsuta2009}.  One of the purposes of this study is to test several modern machine learning algorithms to assess which method is the most accurate in this setting and to compare the performance of the more complex methods to the performance of LDA that was used in the original paper \cite{BrophyEtAl2014}.  The current study will also improve upon the work done in Brophy \cite{BrophyEtAl2014} by expanding the modeling framework to fit two different levels:  1) a tribe classification model and 2) a species classification model conditional on tribe.  This second step was not employed for the extant teeth in Brophy \cite{BrophyEtAl2014}.  Five different machine learning techniques including LDA, neural networks (NNET), nuclear penalized multinomial regression (NPMR), random forests (RF), and support vector machines (SVM), described in detail in section \ref{SMLA}, were used to estimate these models. The results demonstrate that considerable improvements can be achieved over LDA. 

The remainder of this manuscript gives a description of the methods in section \ref{methods} followed by the results of the explorations in section \ref{results}. The paper then concludes with a summary of the results and a discussion of and recommendations for future work in section \ref{conclusions}.

\section{Data} 

\begin{table}
\caption{\label{Data}The number of observations in each tribe/species/tooth grouping}
\centering
\begin{tabular}{cc|rrrrrr}
& Species & LM1 & LM2 & LM3 & UM1 & UM2 & UM3 \\ 
  \hline
Alcelaphini &{\it Damaliscus dorcas} &  30 &  30 &  31 &  29 &  30 &  30 \\ 
 &{\it Alcelaphus buselaphus} &  15 &  17 &  15 &  15 &  15 &  15 \\ 
  &{\it Connochaetes gnou} &  12 &  12 &  12 &  12 &  13 &  12 \\ 
  &{\it Connochaetes taurinus} &   9 &   9 &   9 &   9 &   8 &  10 \\ 
  \hline
 Antilopini & {\it Antidorcas marsupialis} &   9 &   9 &   9 &   7 &   8 &   7 \\ 
 \hline
Tragelaphini  &{\it Taurotragus oryx} &  12 &  15 &  14 &  15 &  15 &  29 \\ 
 &{\it Tragelaphus strepsiceros} &   8 &  11 &  11 &  10 &  11 &  14 \\ 
  &{\it Tragelaphus scriptus} &   6 &  11 &   9 &   9 &  11 &  15 \\ 
  \hline
Bovini   &{\it Syncerus caffer} &  15 &  15 &  15 &  15 &  15 &  30 \\ 
\hline
  Neotragini &{\it Raphicerus campestris} &  12 &  15 &  15 &  15 &  15 &  29 \\ 
  &{\it Oreotragus oreotragus} &  15 &  14 &  15 &  15 &  15 &  24 \\ 
  &{\it Pelea capreolus} &  22 &  29 &  31 &  31 &  30 &  30 \\ 
  &{\it Ourebia ourebi} &  15 &  15 &  15 &  15 &  15 &  27 \\ 
  \hline
  Hippotragini &{\it Hippotragus niger} &  30 &  30 &  30 &  28 &  28 &  30 \\ 
  &{\it Hippotragus equinus} &  24 &  27 &  25 &  29 &  31 &  30 \\ 
  &{\it Oryx gazella} &  27 &  30 &  30 &  27 &  30 &  30 \\  
\hline
Reduncini  &{\it Redunca arundinum} &  15 &  23 &  15 &  31 &  31 &  29 \\ 
  &{\it Redunca fulvorufula} &  15 &  24 &  15 &  30 &  31 &  15 \\ 
  &{\it Kobus leche} &  15 &  30 &  15 &  32 &  32 &  25 \\ 
  &{\it Kobus ellipsiprymnus} &  15 &  15 &  15 &  15 &  15 &  15 \\ 
   \hline
& Total & 321 & 381 & 346 & 389 & 399 & 446
 \end{tabular}
\end{table}

The data consisted of a total of 2282 modern bovid teeth from six different tooth types: mandibular (lower) molars 1, 2, and 3 (LM1, LM2, LM3) and maxillary (upper) molars 1, 2, and 3 (UM1, UM2, UM3).  The reference data set is based on modern bovids of known taxonomic identify. While isolated teeth are difficult to identify, the modern bovid images were obtained from crania with associated teeth and horns, another useful way of identifying bovids.  These teeth include bovids from 7 different tribes (Alcelaphini, Antilopini, Tragelaphini, Bovini, Neotragini, Hippotragini, and Reduncini) and 20 different species. The number of species within a tribe varied from 1 (Antilopini and Bovini) to 4 (Alcelaphini, Neotragini, and Reduncini). When constructing these models, each tooth classification (e.g. LM1) was considered separately.  The number of observed teeth for each tooth type varied from a low of 321 observation (LM1) to a high of 446 (UM3).  For a species and a tooth type, the number of observations ranged from 7 (Tragelaphini {\it Tragelaphus scriptus}, LM1) to  32 (Reduncini {\it Kobus leche}, UM1).  Full details of the number of observations in each tribe/species/tooth group can be found in Table \ref{Data}.

\section{Methods}\label{methods}
\subsection{Edge extraction and Elliptical Fourier Analysis}
\begin{figure}
\includegraphics[width=\linewidth]{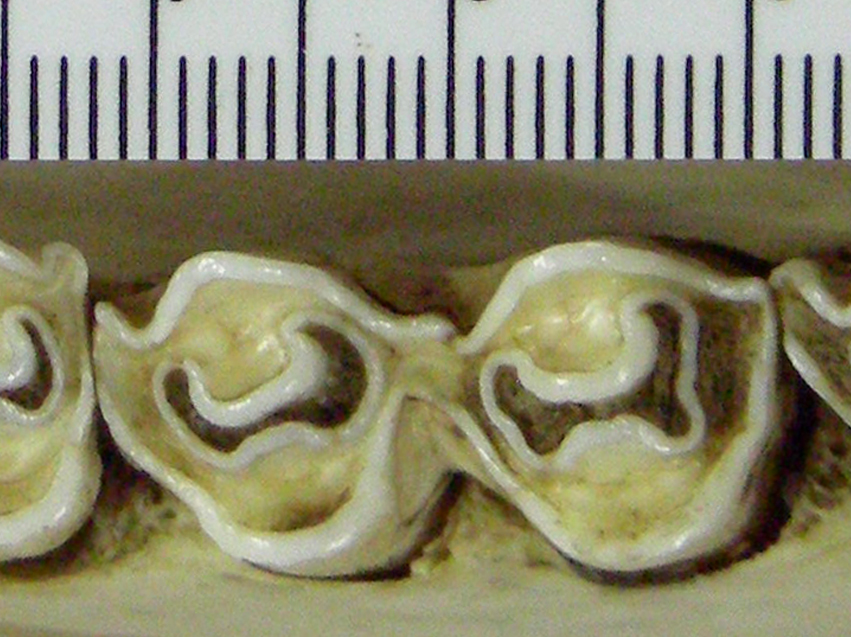}
\caption{Example Tooth}
\label{exampleTooth}
\end{figure}

Brophy \cite{BrophyEtAl2014} captured two-dimensional bounded outlines of bovid teeth using Elliptical Fourier Function Analysis (EFFA) \cite{Lestrel1989,Wolfe1999}, a curve fitting function particularly suited for defining bounded outline data.  Sixty points were manually placed around each tooth (e.g. Figure \ref{exampleTooth}) according to a template in the program MLmetrics \cite{Wolfe1999}, a digitizing program designed specifically to work with EFFA. The (x,y) coordinates of the outline were imported into EFFA (EFF23 v. 4) where the harmonics and amplitudes were generated \cite{Lestrel1989,Wolfe1999}. EFFA approximates each tooth as a sum of ellipses. Formally, this parametric function is as follows: 
$$
x = f(u) = A_0 + \sum_{j=1}^H a_j cos(ju) + \sum_{j=1}^{H} b_j sin(ju)
$$

$$
y = g(u) = C_0 + \sum_{j=1}^H c_j cos(ju) + \sum_{j=1}^{H} d_j sin(ju)
$$ 
where $H$ is the number of harmonics used, $A_0$ and $C_0$ are constants, and $a_j$, $b_j$, $c_j$, and $d_j$ are the amplitudes associated with the $j$-th harmonic and $j = 1, 2, \cdots, H$.  The amplitudes (i.e. $a_j$, $b_j$, $c_j$, and $d_j$)  for each tooth type (e.g. LM1) were compared across species in the same bovid tribe. Next, principal component analyses (PCA) on the covariance matrix of the amplitudes were calculated and used as additional features when performing LDA \cite{Fisher1936,TibshiraniHasiteFriedman}.

\subsection{Supervised Machine Learning Algorithms}\label{SMLA}
This research investigated five machine learning techniques in order to determine which approach best classifies the tribe and species of modern bovid teeth using the output of EFFA as features of the model. These techniques include LDA, neural networks (NNET), nuclear penalized multinomial regression (NPMR), random forests (RF), and support vector machines (SVM). This paper assesses which complex machine learning algorithm results in the highest predictive accuracy in tooth identification. 


Let $T_i$ and $S_i$ be the true tribe and species of the $i$-th tooth, respectively, with $T_i \in (1,2,\cdots,K_{T})$ and $S_i \in (1,2,\cdots,K_{S})$. Further, let ${\bf T} = (T_1, T_2,\cdots, T_n)$ and ${\bf S} = (S_1, S_2,\cdots, S_n)$ so that both ${\bf T}$ and ${\bf S}$ are vectors of length $n$.  

Also, for the $i$-th tooth let the vector 
$$
{\bf A}_i = (a_{i1},b_{i1},c_{i1},d_{i1},a_{i2},b_{i2}, \cdots, a_{iH},b_{iH},c_{iH},d_{iH})
$$, 
which will have length $4H$ where $H$ is the number of harmonics.  Here $H=15$ is used.  Let matrix ${\it A} = ({\bf A}_1, {\bf A}_2, \cdots, {\bf A}_n)'$, which will contain $n$ rows and $4H$ columns.   Finally, let ${\it \Sigma}_{A} = cov(A)$ and using spectral decomposition we have ${\it \Sigma}_{A} = {\it P}'{\it \Lambda}^{\frac{1}{2}} {\it \Lambda}^{\frac{1}{2}} {\it P}$ where ${\it \Lambda}$ is a diagonal matrix whose elements are the eigenvalues of ${\it \Sigma}_{A}$ and ${\it P}$ is a a matrix of eigenvectors of ${\it \Sigma}_{A}$ \cite{JohnsonAndWichern2002}.  Now define ${\it Y} = {\it A} {\it \Gamma}$ where ${\it Y}$ is a $n$ by $4H$ matrix of the coordinates rotated by the principal components and ${\it \Gamma} = {\it \Lambda}^{\frac{1}{2}}{\it P}$.  The matrix ${\it A}$ and the first $p$ columns of ${\it Y}$ are used as potential predictors in each of the five machine learning algorithms.  The augmented matrix is denoted ${\it X} = ({\it A}:{\it Y})$ where ${\it X}$ is the augmented matrix combining ${\it A}$ and ${\it Y}$ that contains $n$ rows and $4H + p$ columns where $p \le 4H$.  

The teeth are classified at two different taxonomic levels: 1) tribe and 2) species.  In order to achieve this, two levels of modeling were performed. In the first level, the model was built to classify the tribe of an observation given the features derived from EFFA of the tooth and the principal components. The model is as follows: $p^{(T)}_{ik} = P(T_{i}=k|{\bf X}), i = 1, \cdots, n$.  These probabilities were estimated using the five approaches (LDA, NPMR, RF, SVM, NNET).  Following this, a series of models were fit for species classification conditional on each tribe.  Formally, $p^{(S|T=k)}_{ig} = P(S_i = g|T_i = k,{\bf X})$ for $i = 1, \cdots, n$.  This conditional probability is then used to compute $p^{(S,T)}_{igk} = P(S_i = g, T_i = k|{\bf X}) = P(S_i = g|T_i = k ,{\bf X})P(T_i = k|{\bf X})$.  One convenient property of structuring the models in this manner is that 
$$
\sum_{g\in S(t)}P(S_i=g, T_i=k|{\bf X}) = P(T_i=k|{\bf X})
$$
where $S(t)$ is the set of species that fall into tribe $T=k$.  This means that when estimating the probability that a particular tooth belongs to a particular tribe, the implied probability that a tooth belongs to a particular tribe that can be calculated by summing across all of the species that are nested within a particular tribe and this probability will be consistent with probability calculated from the tribe model alone.  This entire modeling process was repeated separately for the 6 different teeth considered (e.g. LM1, LM2, etc.).

Note that in the following subsection each method is described in terms of the tribe classification model for illustration purposes. The methods are easily modified to classify species within a given tribe.  

\subsubsection{Linear Discriminant Analysis}
Linear discriminant analysis \cite{Fisher1936,TibshiraniHasiteFriedman} assumes that the distributions of the features within each specific tribe follow multivariate Gaussian densities all with common covariance matrix, ${\it \Sigma}_{LD}$, across all of the $K_{T}$ tribes.  If we then denote the conditional density of $X|T$ as $f_k(x) = P(X = x|T = k)$ and let $\pi_{k}$ be the prior probability of belonging to tribe $k$ then the discriminant function is defined as follows: 

$$
\delta_{k}(x) = x^{\top}{\it \Sigma}^{-1}\mu_{k} - \frac{1}{2}\mu_{k}^{\top} + log(\pi_{k})
$$. 

The parameters in the discriminant function are unknown and must be estimated from the data and can be computed as
$$
\hat{\pi}_{k} = \frac{n_{k}}{n}
$$

$$
\hat{\mu}_{k} = \sum_{i=1}^{n_{k}}\frac{x_i}{n_k}
$$

$$
\hat{{\it \Sigma}} = \sum_{k=1}^{K_{T}}\sum_{i=1}^{n_{k}}\frac{(x_i - \hat{\mu}_k)(x_i - \hat{\mu}_k)^{\top}}{n-n_k}
$$

where $n_k$ is the number of observation in the $k$-th tribe and $n = \sum_{k=1}^{K_{T}}n_k$.  Each observation is then classified as follows:  $\hat{T}_i = argmax_{k}\delta_{k}(x_i)$.  

LDA was implemented here using the ``lda" function from the MASS package \cite{LDA} in R \cite{R}.

\subsubsection{Nuclear Penalized Multinomial Regression}
Multinomial regression assumes that the distribution of tribes follow a multinomial distribution conditional on predictor variables, which in this setting are the amplitudes of the harmonics and principal components of the amplitudes.   
 The multinomial model is specified as a series of logit transformations such as

$$
log\left(\frac{P(T = 1|X= x)}{P(T= K_{T}|X = x)}\right) = \alpha_1 + {\bf \beta}_{1}X
$$

$$
log\left(\frac{P(T = 2|X= x)}{P(T= K_{T}|X = x)}\right) = \alpha_2 + {\bf \beta}_{2}X
$$
{\centering$\vdots$}
$$
log\left(\frac{P(T = K_T-1|X = x)}{P(T = K_{T}|X = x)}\right) = \alpha_{K_{T}-1} + {\bf \beta}_{K_T-1}X
$$. 

In this setting, each ${\bf \beta}_k$ is a vector of length $4H+p$, and there are a total of $K_{T}-1$ of these vectors, which can then be combined into a matrix, which will be referred to as ${ \bf B}$ with dimension $4H+p$ by $K_{T}-1$. 

These equations can be used to define the log-likelihood $\ell(\alpha, { \bf B}; {\it X}, {\bf T})$, and the maximum likelihood estimate of the parameters is found by  
$$
\max\limits_{\alpha\in \mathbb{R}^{K_{T}-1}, {\bf B}\in \mathbb{R}^{4H+p \times K_{T}-1}} \ell(\alpha, {\bf B}; {\it X}, {\bf T})$$.

In nuclear penalized multinomial regression (NPMR), the same set-up is assumed as in multinomial regression, however, a penalty term is added to the likelihood. Coefficient estimates are found by maximizing the likelihood subject to this constraint using the nuclear norm on the matrix ${\bf B}$.  Specifically, estimates are found by maximizing the following expression:  

$$
\max\limits_{\alpha\in \mathbb{R}^{K_{T}-1}, {\bf B}\in \mathbb{R}^{4H+p \times K_{T}-1}} \ell(\alpha, {\bf B}; {\it X}, {\bf T}) - \lambda ||{\bf B}||_{\star}
$$ 

where 
$$
||{\bf B}||_{\star} = \sum_{r=1}^{rk({\bf B})} \sigma_r
$$
and $\sigma_r$ are the singular values of ${\bf B}$.  Optimal values of $\lambda$ were found using cross validation, with 15 powers of $e$ considered ranging from -3 to 2 for the tribe models and 25 powers of $e$ ranging from -5 to 3 for the species model.  In the tribe models, cross validation identified values of $\lambda$ were found across the entire range of values tried depending on the tooth type and fold. Results of the species models were similar in that chosen values of $\lambda$ came from the entire range of values considered. NPMR was implemented in this study using the ``npmr" package \cite{Powers2016} in R \cite{R}.  

\subsubsection{Random Forests}
Random Forests \cite{Breiman2001} consist of a collection of classification trees where each tree ``votes'' on the correct tribe for observation $i$ where each tree is built from a bootstrapped \cite{Efron1979} sample of the original data. 

In the tribe model, the model attempts to classify each observation into one of $K_T$ categories.  A tree model involves partitioning the feature space into $M$ regions $R_{1}, R_{2}, \cdots, R_{M}$.

$$
\hat{p}_{mk} = \frac{1}{n_m}\sum_{x_{i}\in R_m} I(T_i = k)
$$
where $m$ is the node index of region $R_m$ which contains $N_m$ observations such that $\sum_{m=1}^{M}n_{m} = n$.  That is, the model will classify observations that fall into node $m$ as the majority class in node $m$, and uses the Gini index as the measure of node impurity 

$$
Gini_{m}(\tau) = \sum_{k\ne k'}\hat{p}_{mk}\hat{p}_{mk'} = \sum_{k=1}^{K_{T}}\hat{p}_{mk}(1-\hat{p}_{mk})
$$
for tree $\tau$.  

Further, at each potential split in all trees, only a subset of the variables are considered as possible splits which further reduces the correlation between the trees beyond bootstrapping alone.  Using cross validation the number of variables to consider for a split at each step was evaluated.  Values considered for the number of splits ranged from 10 to 50 in increments of 10.  This was done for both the tribe and species models and in all cases 2000 trees were used in each random forest.  In the tribe models, the number of variables considered at each step was most often chosen to be 10 and never larger than 30.  In the species models, the number of variables considered at each split predominantly ranged from 10 to 30, though values of 40 and 50 were observed in a few cases.

Here, the random forest algorithm was implemented in R \cite{R} using the package \texttt{randomForest} \cite{randomForest}.



\subsubsection{Support Vector Machines}
Support vector machines (SVM) \cite{TibshiraniHasiteFriedman} perform classification by constructing linear boundaries in a transformed version of the feature space.

Formally, if the model is seeking to classify two different tribes, in the case when the two groups are completely separable, a support vector classifier seeks a hyperplane of the form 
$$
\{x : f(x) = x^{\top}{\bf \beta} +\beta_{0} = 0\}
$$
where $||{\bf \beta}||=1$.  The model then classifies an observation to the two different classes based on the sign of $f(x)$.  This amounts to finding $max_{{\bf \beta},\beta_0,||{\bf \beta}||=1} C$ where $C=\frac{1}{||{\bf \beta}||}$ and with the imposed constraint that $y_i (x^{\top}_{i}{\bf \beta} +\beta_0)\ge C$, for $i = 1, 2, \cdots, n$ where $y_{i} \in \{-1, 1\}$ depending on the tribe that the $i$-th observation belongs to. 

When groups (i.e. tribes) are not completely separable, the constraints must be relaxed.   This method is achieved by defining a set of slack variables ${\bf \xi} = (\xi_{1},\xi_{2},\cdots,\xi_{n})$ and again maximizing $C$. The constraint now is modified to be $y_{i} (x^{\top}_{i}{\bf \beta} + \beta_0) \ge C(1-\xi_{i})$ where $\xi_{i} \ge 0$ for all $i$ and the sum of all $\xi_{i}$ is bounded by a constant.  

This method works well when the boundaries between two classes is linear.  However, the support vector classifier can be extended by considering a larger feature space through the use of specific functions called kernels.  By choosing a linear kernel, the model reduces to the support vector classifier.  Other possible kernels include a polynomial kernel and a sigmoid kernel, for example.  In this setting the kernel that was found to work best was the radial kernel, which was used here when fitting the SVM models.  Values for the tuning parameter of the radial kernel that were considered were powers of 2 with powers ranging from -10 to 3 in increments of 0.5.  Cost parameters considered were powers of ten with the powers ranging from -5 to 5 in increments of 1.  Through cross-validation, the value of the cost parameter was always chosen to be either 1, 10 or 100.  Values of the tuning parameters for the radial kernel were found to be mostly between values of $2^{-10}$ and $2^{-8}$ and never larger than $2^{-6}$.

While SVM methods are used to discriminate between two different groups, this approach can be extended when the number of classes is greater than 2 (here there are 7 tribes and 20 species.)  By using  the one-against-one approach where all pairwise classifications are considered, an observation is classified based on a voting scheme where the final classification is based on which class wins more of the pairwise comparisons.  

Here we implemented the SVM algorithm in R using the package \texttt{e1071} \cite{e1071}.

\subsubsection{Neural Networks}
Neural networks \cite{TibshiraniHasiteFriedman} can be thought of as two-stage classification models. The first stage transforms the input features in matrix ${\it X}$ into a matrix of derived features ${\it Z}$ with dimension $n$ by $Q$.  Then each tribe is modeled as a function of the matrix ${\it Z}$, where ${\it Z}$ is often referred to as the hidden layer in these models. Formally, 
$$ 
Z_q = \sigma(\zeta_{0q}+\zeta^{T}_{q} {\bf X}), q = 1,\cdots, Q 
$$
$$
U_k = \beta_{0k} + {\bf \beta}^{\top}_{k}{\it Z}, k = 1,\cdots, K_{T}
$$
$$
f_{k}(X) = g_{k}(U), t = 1,\cdots, K_{T} 
$$

where $\sigma$ is the activation function commonly chosen to be $\sigma(\eta) = \frac{1}{1+e^{-\eta}}$.  The function $g$ allows for a transformation from the output $U_k$.  In classification problems this is usually chosen to be $g_{k}(U) = \frac{e^{U_{k}}}{\sum_{j=1}^{K_{T}}e^{U_j}}$ which is the same transformation function that is used in multinomial regression problems.  Tuning parameters for NNET were found using cross validation.  Values of 2, 4, 6, and 8 were considered for the size of the hidden layer and powers of 2 between -10 and 2 (15 evenly spaced values) were considered for the decay parameter of the model.  In the tribe models the optimal size was 8 in all teeth across all folds and the decay parameters were found to be optimal on the upper end of the values considered (i.e. 1.587 and 4).  In the species models the size was most often optimal at 8, however, sizes of 6, 4, and 2 were also used in some settings.  Values of the decay parameter for the species models varied much more widely than the the tribe models and optimal parameters ranged across all values considered. 

Neural nets were implemented using the \texttt{nnet} \cite{nnet} package in R \cite{R}, which fits a neural network with a single hidden layer.


\section{Results}\label{results}
A 6-fold cross validation was used to evaluate the accuracy of each classification method of each tooth type for both tribe and species.  It should be noted that in the methods that require tuning parameters, these were found using an inner cross validation where only the 5 of 6 folds that comprise the trained data set in each step of the cross validation were used to tune the model.  We evaluated each method using log loss and classification accuracy.  Log loss for the tribe was computed as follows: 

$$
logLoss_{Tribe} = \sum_{i=1}^{n}\sum_{k=1}^{K_{T}} I(T_{i}=k) log(\hat{p}^{(T)}_{ik})
$$

where $\hat{{\bf p}}^{(T)}_{i}$ is the vector of length $K_{T}$ containing the predicted probabilities of each of the $K_{T}$ tribes for the $i$-th observation and $I(T_{i}=k)$ is an indicator function which is equal to 1 if $T_{i}=k$ and 0 otherwise.  

Log loss for the species models was computed in a similar fashion as follows:  
$$
logLoss_{Species} = \sum_{i=1}^{n}\sum_{g=1}^{K_{S}} I(S_{i}=g) log(\hat{p}^{(S,T)}_{ig})
$$

where $\hat{{\bf p}}^{(S,T)}_{i}$ is the vector of length $K_{S}$ containing the predicted probabilities of each of the $K_{S}$ species for the $i$-th observation and $I(S_{i}=g)$ is an indicator function which is equal to 1 if $S_{i}=g$ and 0 otherwise.  

Log loss was chosen due to the fact that this function heavily penalizes a method for being over-confident and wrong. In this setting, we prefer a classification method that avoids being incorrectly over-confident in classification.  

Further, accuracy was used as another way to evaluate each model.  For each probability vector prediction for observation $i$, a tooth was classified as belonging to the tribe or species with the highest predicted probability in the vector.  This result was then compared to the actual tribe/species of the observation. 

$$
Accuracy_{Tribe} = \frac{\sum_{i=1}^{n}{I(T_{i} = \hat{T}_{i})}}{n}
$$
where $\hat{T}_{i} = \{k : \underset{\text{k}}{max}$ $\hat{p}^{(T)}_{ik}\}$.  

$$
Accuracy_{Species} = \frac{\sum_{i=1}^{n}{I(S_{i} = \hat{S}_{i})}}{n}
$$
where $\hat{S}_{i} = \{g : \underset{\text{g}}{max}$ $\hat{p}^{(S,T)}_{igk}\}$. 

This is simply the percentage of teeth that are classified to the correct tribe or species given the classification rule that classifies an observation to the most likely category (i.e. the category with the largest predicted probability).  

\subsection{Log Loss Results}
\begin{figure}[!htb]
\centering
\includegraphics[width=.8\linewidth, height=4in]{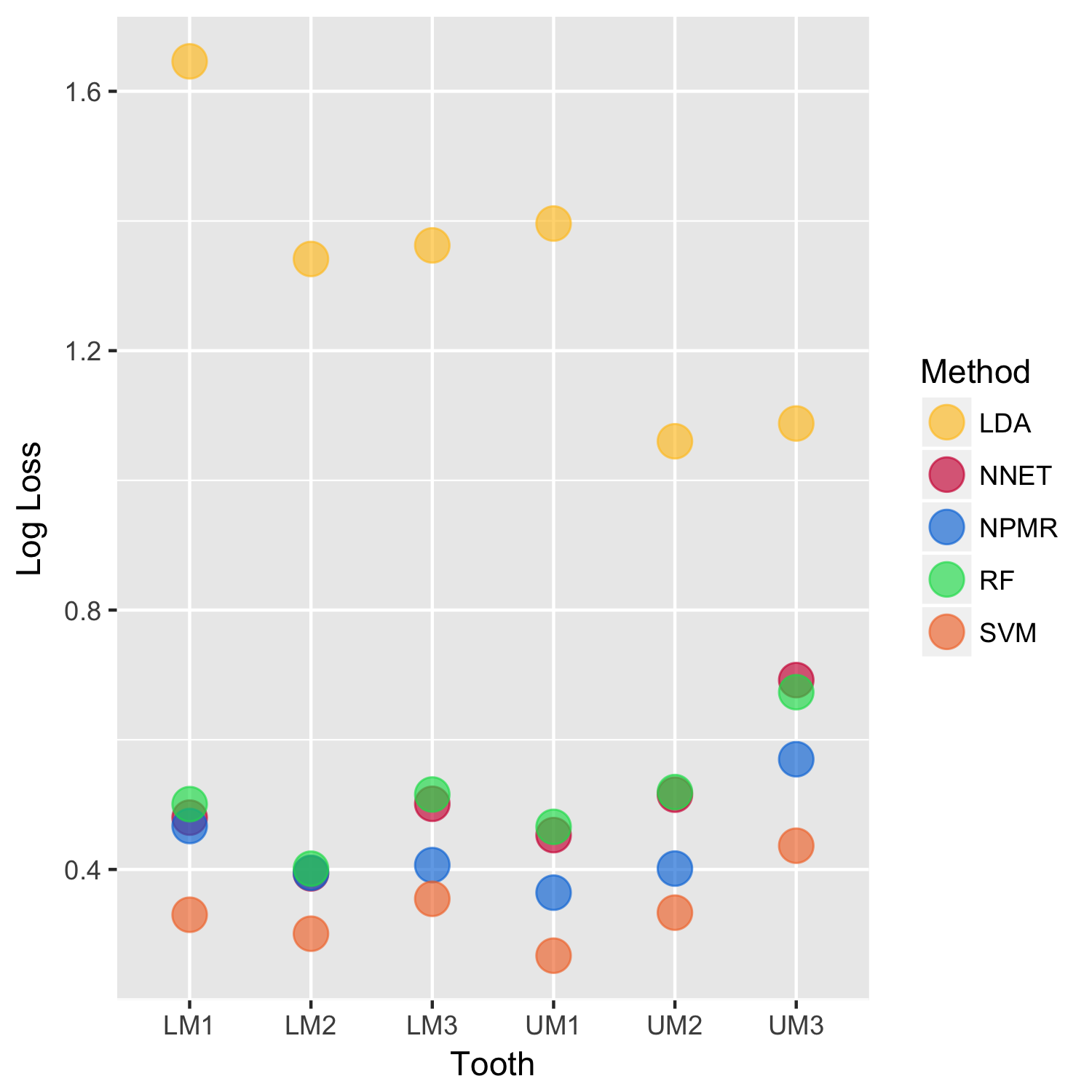}  
\caption{Tribe Prediction Model}
\label{figTribeLog}
\end{figure}

In terms of log loss when predicting tribe, which can be seen in the left five columns of figure \ref{tabLogLoss}, SVM was the best model in all six categories for tribe prediction.  Log loss values of SVM ranged from 0.27 through 0.44.  NPMR performed the second best across all six teeth with followed by either NNET and RF depending on the tooth.  LDA was uniformly the worst across all six teeth with log loss values ranging from 1.06 through 1.65.  Full log loss results for tribe classification can be seen in the left 5 columns of table \ref{tabLogLoss}, and these results are illustrated in figure \ref{figTribeLog}.    

\begin{figure}[!htb]

        \centering
        \includegraphics[width=.8\linewidth, height=4in]{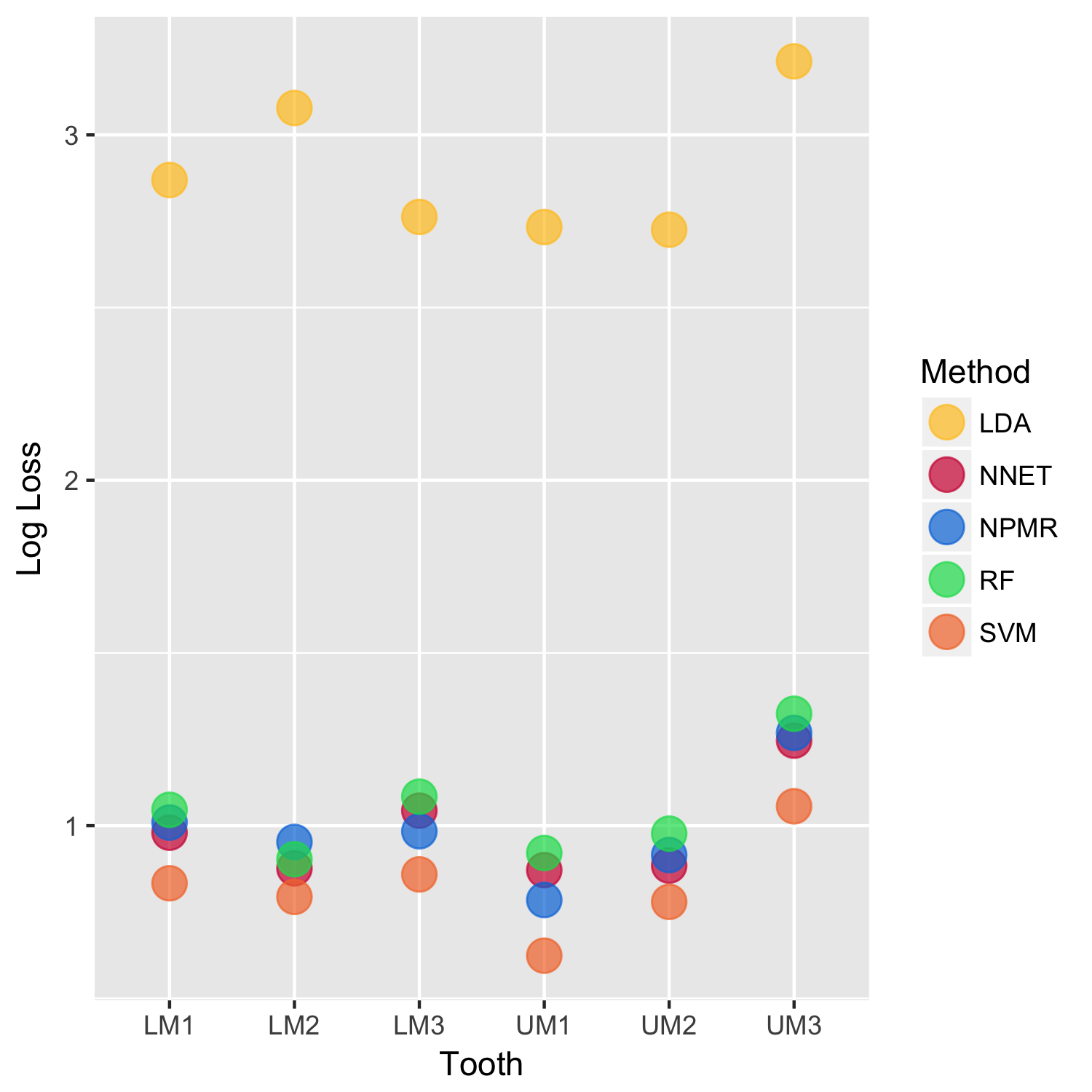}  
\caption{Species Prediction Model}
\label{figSpeciesLog}
\end{figure}

The species models results for log loss are similar to that for predicting tribe with SVM performing the best across all teeth with log loss values ranging from 0.62 through 1.06.  However, when predicting species there is not a clear second best model with NNET, NPMR, and RF all performing similarly.  Again LDA is the worst model in terms of predicting species with log loss values ranging from 2.73 through 3.21.  Full log loss results for tribe classification can be seen in the right 5 columns of table \ref{tabLogLoss}, and these results are illustrated in figure \ref{figSpeciesLog}.

\begin{table}
\caption{\label{tabLogLoss} Log loss values for tribe and species classification for each tooth and machine learning method considered}
\centering
\begin{tabular}{|c|ccccc|ccccc|}
\hline
& \multicolumn{5}{|c|}{Tribe} &  \multicolumn{5}{|c|}{Species}  \\
  \hline
 & NNET & NPMR & RF & SVM & LDA & NNET & NPMR & RF & SVM & LDA \\ 
  \hline
LM1 & 0.48 & 0.47 & 0.50 & {\bf 0.33} & 1.65 & 0.98 & 1.01 & 1.05 & {\bf 0.83} & 2.87 \\ 
  LM2 & 0.39 & 0.39 & 0.40 & {\bf 0.30} & 1.34 & 0.88 & 0.95 & 0.90 & {\bf 0.79} & 3.08 \\ 
  LM3 & 0.50 & 0.41 & 0.52 & {\bf 0.35} & 1.36 & 1.04 & 0.98 & 1.08 & {\bf 0.86} & 2.76 \\ 
  \hline
  UM1 & 0.45 & 0.36 & 0.47 & {\bf 0.27} & 1.40 & 0.87 & 0.78 & 0.92 & {\bf 0.62} & 2.73 \\ 
  UM2 & 0.52 & 0.40 & 0.52 & {\bf 0.33} & 1.06 & 0.88 & 0.91 & 0.98 & {\bf 0.78} & 2.73 \\ 
  UM3 & 0.69 & 0.57 & 0.67 & {\bf 0.44} & 1.09 & 1.25 & 1.27 & 1.32 & {\bf 1.06} & 3.21 \\ 
   \hline
\end{tabular}
\end{table}

\subsection{Accuracy Results}
\begin{figure}[!htb]
\centering
        \includegraphics[width=.8\linewidth, height=4in]{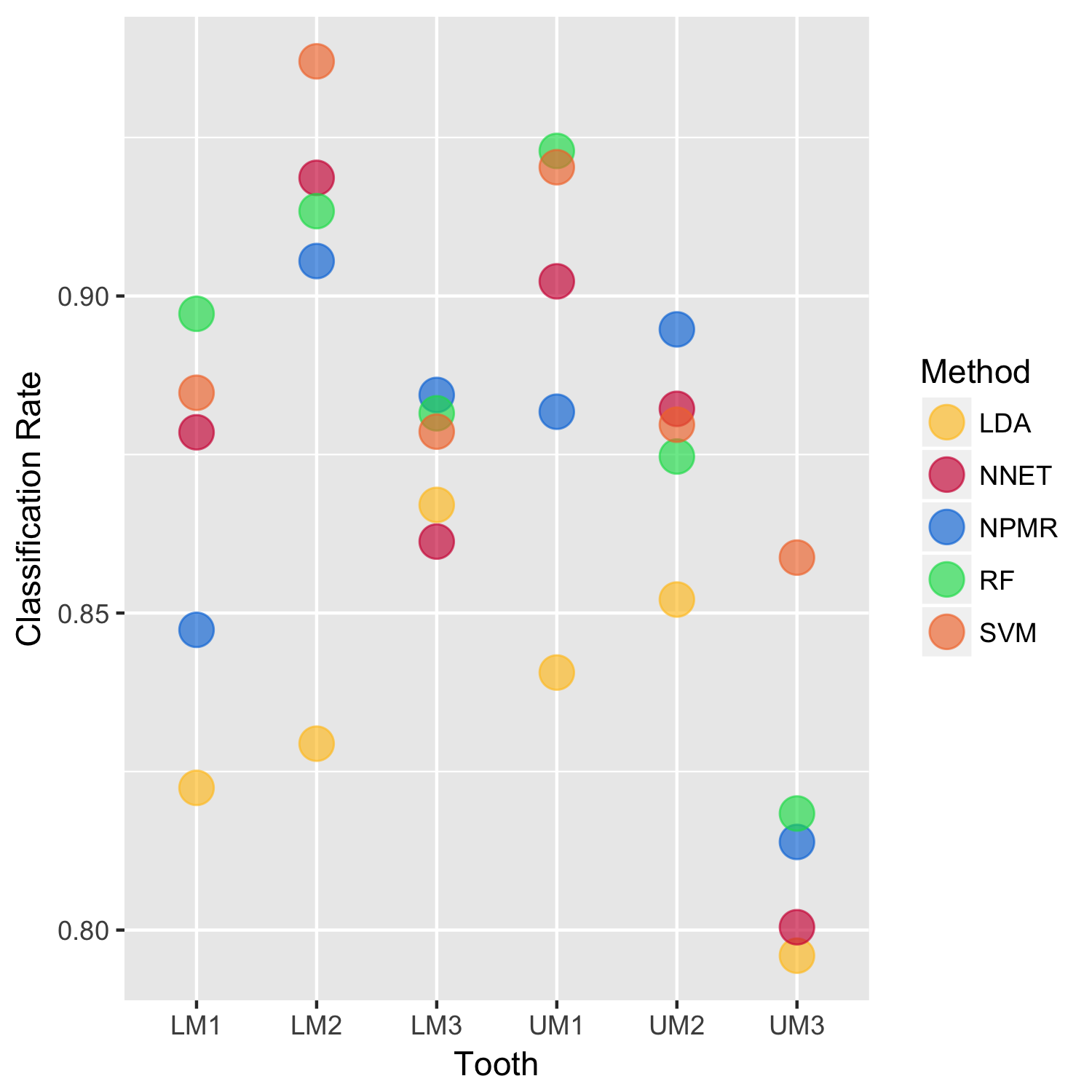}  
\caption{Tribe Prediction Model Accuracy Results}
\label{figTribeAcc}
\end{figure}

In terms of accuracy for the tribe models (see table \ref{tabAccuracy} and figure \ref{figTribeAcc}), there was no clear method that performed the best in this setting.  Of the six teeth tested, NPMR, RF, and SVM performed the best on 2 teeth each.  For classifying tribe, most methods perform very well with only one method and tooth combination classifying under 80\% (LDA and UM3, 79.6\%).  While all of the methods perform fairly well, the only method to exceed 85\% accuracy across all six teeth was SVM. 

\begin{figure}[!htb]

        \centering
        \includegraphics[width=.8\linewidth, height=4in]{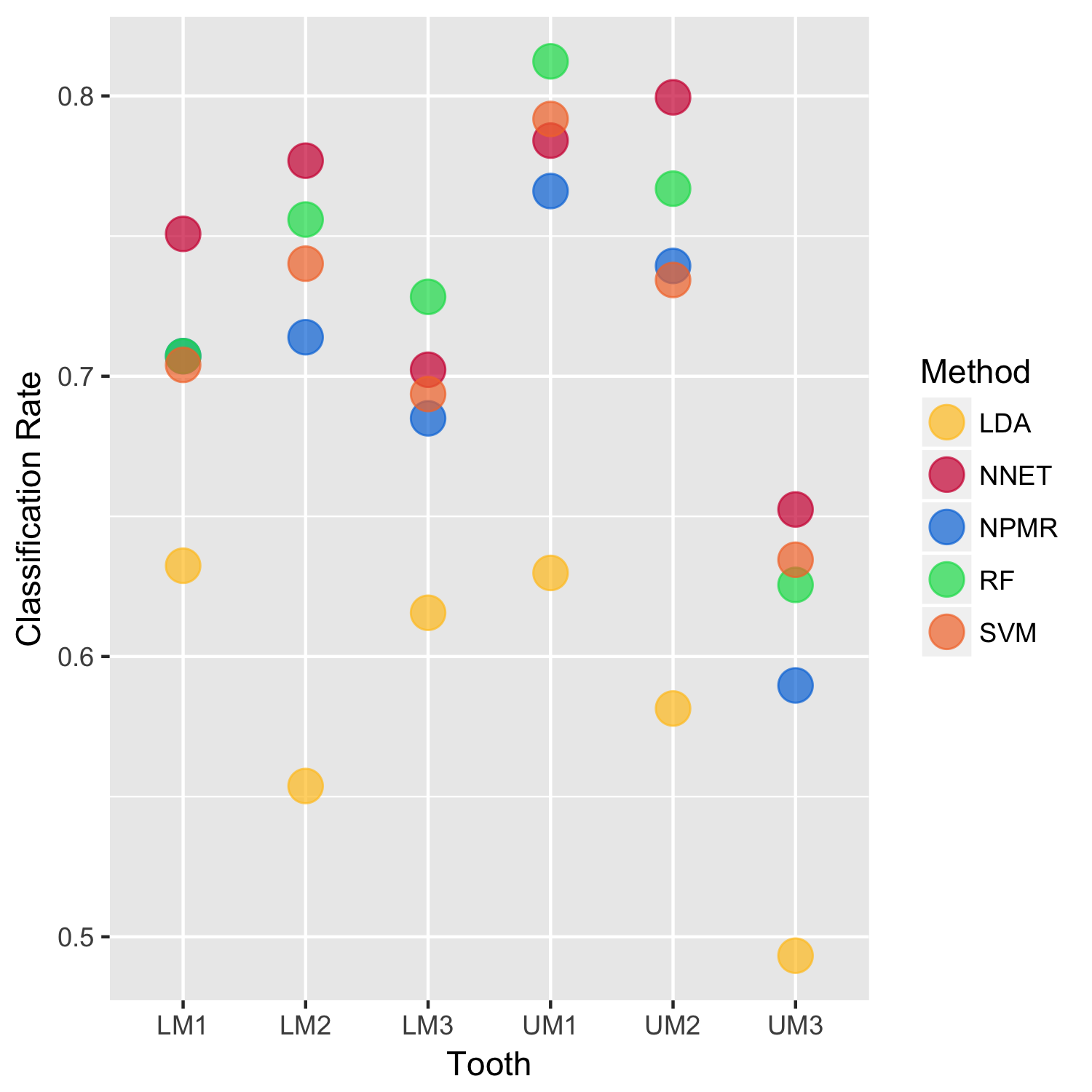}  
\caption{Species Prediction Model Accuracy Results}
\label{figSpeciesAcc}
\end{figure}

Oddly enough, in the species modeling (results in figure \ref{figSpeciesAcc}), in four of the six teeth the best performing method is NNET even though it was never the best in the tribe modeling.  In the other two teeth where NNET was not the best, RF outperformed the other methods.  Overall, the species had a lower correct classification rate than the tribe models, which is to be expected. Finally, note that LDA is uniformly the worst method in terms of accuracy for both the tribe and species models.



\begin{table}
\caption{\label{tabAccuracy}Rate of accurate classification for each tooth and machine learning method considered.}
\begin{tabular}{|c|ccccc|ccccc|}
\hline
& \multicolumn{5}{|c|}{Tribe} &  \multicolumn{5}{|c|}{Species}  \\
  \hline
 & NNET & NPMR & RF & SVM & LDA & NNET & NPMR & RF & SVM & LDA \\ 
  \hline
LM1 & 0.879 & 0.847 & {\bf 0.897} & 0.885 & 0.822 &{\bf 0.751} & 0.707 & 0.707 & 0.704 & 0.632 \\ 
  LM2 & 0.919 & 0.906 & 0.913 & {\bf 0.937} & 0.829 & {\bf 0.777} & 0.714 & 0.756 & 0.740 & 0.554 \\ 
  LM3 & 0.861 & {\bf 0.884} & 0.882 & 0.879 & 0.867 & 0.702 & 0.685 & {\bf 0.728} & 0.694 & 0.616 \\ 
  \hline
  UM1 & 0.902 & 0.882 & {\bf 0.923} & 0.920 & 0.841 & 0.784 & 0.766 & {\bf 0.812} & 0.792 & 0.630 \\ 
  UM2 & 0.882 & {\bf 0.895} & 0.875 & 0.880 & 0.852 & {\bf 0.799} & 0.739 & 0.767 & 0.734 & 0.581 \\ 
  UM3 & 0.800 & 0.814 & 0.818 & {\bf 0.859} & 0.796 &{\bf 0.652} & 0.590 & 0.626 & 0.635 & 0.493 \\ 
   \hline
\end{tabular}
\end{table}

\section{Discussion and Conclusions}\label{conclusions}
Reconstructing past environments associated with hominins is an important aspect of understanding early human behavior. Therefore, correctly classifying the bovid teeth recovered from hominin sites that are used in paleoenvironmental reconstruction is an essential  task. 

One issue that needs to be addressed is the extraction of the outline of the occlusal surface of the teeth. Previously, this has been performed manually by an expert (e.g. JKB).  However, the process of edge extraction is tedious, time consuming, and impossible to perform in a practical manner with the quantity of data that we plan to collect moving forward.  Therefore, along with this work, we have been exploring efficient methods for extracting the occlusal surface of these specimens.  One method that we have explored is the use of crowd sourcing through Amazon's Mechanical Turk platform for the edge extraction, and our initial work in this area is highly encouraging.

This paper expands on the work in Brophy  \cite{BrophyEtAl2014} by examining  more flexible machine learning algorithms beyond LDA and refines the assessment of classification accuracy by using external cross validation.  Here, five machine learning techniques, including LDA, are compared based on their performance in classifying the tribe and species of modern extant bovid molars with the amplitudes produced by elliptical Fourier analysis.  This work demonstrates that classification accuracy can be increased in terms of both log-loss and misclassification rate by using support vector machines and random forests, both of which performed similarly and slightly outperformed nuclear penalized multinomial regression. Support Vector Machines and RF performed much better than NNET and LDA.  Moving forward, when bovid teeth are digitized from new sites, these new techniques for classification of fossilized teeth will be utilized.  

A future goal is to expand the training data set to include a much larger sample of extant bovid molars. Further, two particular fossil sites that are of immediate interest in applying these classification techniques to are Swartkrans and Coopers Cave in South Africa; subsequently, the research would expand to other South African sites including Makapansgat, Sterkfontein, Equus Cave, and Nelson Bay Cave, to name a few. The ultimate goal is to be able to discuss accurately  broad environmental changes in southern Africa in the Plio-Pleistocene and examine if fluctuations in the paleoenvironment correlate with hominin evolution.

\newpage
\bibliographystyle{tfs}
\bibliography{brophyTeeth}
\end{document}